\documentstyle[12pt]{article}
\begin{document}
\begin{titlepage}
\vspace*{2cm}
\begin{center}
{\Large\bf
Test of the Equivalence Principle from Neutrino Oscillation
Experiments}\vspace{1cm}\\
R.B. Mann\footnotemark\footnotetext{email: rbm20@amtp.cam.ac.uk
$\quad$ on leave from
Physics Department, University of Waterloo,
Waterloo, Ontario, Canada N2L 3G1}\\
Department of Applied Mathematics and Theoretical Physics \\
University of Cambridge, Silver St., Cambridge CB3 9EW \\
and\\
U. Sarkar\footnotemark\footnotetext{email: utpal@prl.ernet.in}\\
Theory Group, Physical Research Laboratory\\
Ahmedabad 380 009\\
India\\
\today \\ DAMTP/R-95/28
\end{center}

\begin{abstract}
We consider the hypothesis that neutrino oscillation data can be
explained if the gravitational couplings of (massless or
degenerate mass) neutrinos are flavour non--diagonal, in
violation of the equivalence principle.  We analyze the various
neutrino oscillation laboratory experimental data including the
recent LSND observations to constrain the relevant parameter
space.  We find that there is no allowed region of parameter
space which can explain the existing data, implying that the LSND
result cannot be explained by oscillations of degenerate-mass
neutrinos due to equivalence principle violations.
\end{abstract}
\end{titlepage}

\pagebreak

Empirical evidence supporting neutrino-flavour oscillations
continues to mount \cite{anjan}. At present there are four
different solar neutrino experiments \cite{solar}, each using
distinct detection techniques, that consistently find a
discrepancy between the measured  solar $\nu_e$ flux and that
predicted by solar models \cite{solmod}. There are also a number
of experiments on atmospheric neutrinos \cite{atm} which find
that the ratio of the flux of $\nu_\mu$ to $\nu_e$ is
significantly smaller than one would expect from standard
particle physics models \cite{barish}. Most recently, the Liquid
Scintillator Neutrino Detector (LSND) group has recently
announced an excess $\overline{\nu_e}$ events between the energy
range 36 and 60 MeV \cite{LSND}.  If this excess is due to the
$\overline{\nu_\mu} \to \overline{\nu_e}$ oscillation, then it
implies an oscillation probability of $(0.34^{+0.20}_{-0.18} \pm
0.07)$\%.  The distance traversed by the $\overline{\nu_\mu}$
before being detected as a $\overline{\nu_e}$ is about 30
metres.

Mechanisms underlying neutrino oscillation typically assume that
neutrinos have nondegenerate masses, following the original
suggestion by Pontecorvo \cite{pont}. In this scenario the weak
interaction eigenstates of neutrinos are distinct from their
mass eigenstates, thereby permitting oscillations between the
various flavours.

An alternative neutrino oscillation mechanism was proposed more
recently by Gasperini \cite{gasp} (and independently by Halprin
and Leung \cite{halp}), in which neutrino weak interaction
eigenstates are distinct from their gravitational eigenstates.
This mechanism (later referred to as the VEP mechanism \cite{Bah})
does not require neutrinos to have nonzero masses; instead neutrino
oscillations occur in this mechanism due to an assumed flavour
non--diagonal coupling of neutrinos to gravity, in violation of
the equivalence principle.

{}From this viewpoint, neutrino oscillation experiments furnish
us with a test of the equivalence principle. The VEP mechanism
has been explored in a number of papers \cite{Veps} as a
possible explanation of solar neutrino data. A recent analysis
\cite{Bah} has shown that, in the context of a two-flavour
model, there are small allowed regions of parameter space at
both small and large VEP mixing angles which are compatible with
present day solar observations. Extension to a full
three-flavour model indicates that the allowed regions of
parameter space can widen due to mixing with a third flavour
\cite{jonas}.

In the present paper we consider the VEP mechanism in the
context of laboratory searches for neutrino oscillations. We
first show that LSND data itself can be explained by  neutrinos
of degenerate or zero mass with flavour non--diagonal
gravitational couplings. We then carry our analysis further to
include other laboratory experiments \cite{E776,skat,other},
which also constrain the allowed VEP parameter space.  We find
that the combination of these constraints rules out any
violation of the equivalence principle in the $e - \mu$ sector,
implying that gravity couples to $\nu_e$ the same way as
$\nu_\mu$.

As a consequence, in the absence of other physical mechanisms
(such as lepton number violation), previous accelerator data in
conjunction with the  LSND results can only be explained by
assuming neutrinos have differing masses. A repetition of the
above analysis in this case would then imply  a new bound on the
parameter space of the VEP mechanism \cite{gasp}.

In this article we shall not argue in what circumstances the
equivalence principle might be violated. Rather, we take a
phenomenological approach to this problem and try to constrain
the parameter space only from an analysis of the existing data
on neutrino oscillations. For the sake of simplicity we work in
the two generation scenario. At the end we shall comment on the
role other oscillation data plays in constraining VEP.

We turn now to the question of whether or not the LSND results
can be understood solely in the context of the VEP mechanism.
In this mechanism, the gravitational eigenstates
$|\nu_G\rangle=(\nu_{1G}, \nu_{2 G})$ are related to the weak
eigenstates $|\nu_W\rangle=(\nu_{e}, \nu_{\mu})$ by an $SO(2)$
rotation $R(\theta_G)$
\begin{equation}
\vert \nu_W\rangle = R(\theta_G)\vert \nu_G\rangle \label{1}
\end{equation}
where $\theta_G$ is the mixing angle. The gravitational
eigenstates are solutions to the Dirac equation in a
Schwarzchild background.  For a spherically symmetric metric,
choosing the trajectory of the neutrino in the radial direction,
we can write down the diagonal components of the Hamiltonian,
\begin{equation}
H^G_i = -2 |\phi (r)| E_i (1 + f_i)  \label{2}
\end{equation}
which governs the evolution of the neutrinos. Here the $f_i$ are
the flavour dependent gravitational parameters, which determine
the magnitude of the violation of the Weak Equivalence
Principle.  The evolution of the weak eigenstates will be
governed by the equation
\begin{equation}
i {d \over d r} \pmatrix{\nu_e \cr \nu_\mu} =
2 E |\phi(r)| \Delta f \pmatrix{0 & {1 \over 2} \sin 2 \theta_G
\cr {1 \over 2} \sin 2 \theta_G & \cos 2\theta_G}
\pmatrix{ \nu_e \cr \nu_\mu}     \label{3}
\end{equation}
where $|\phi (r)|$ is the Newtonian gravitational potential and
$\Delta f \equiv f_2 - f_1$. If the equivalence principle is not
violated then $f_1 = f_2$.

In this paper we shall be discussing small scale terrestrial
laboratory experiments, for which $\phi(r)$ may taken to be
constant.  Although a natural choice for $\phi$ would be the
earth's gravitational potential ($\sim O(10^{-9})$),
another choice is to consider the potential due to all forms of
distant matter. The dominant contribution is from the local
supercluster which has been estimated to be $3 \times 10^{-5}$
\cite{Kenyon}. For our purposes the choice of $\phi$ is irrelevant, since
to find the allowed parameter space we shall
consider $|\phi| \Delta f$ as the relevant parameter.
Particular limits on $\Delta f$ that arise from given
experiments may be found by substituting the above values for $\phi$.

Consider a beam of muon neutrinos that
traverses a distance of L meters. The
probability that a muon neutrino
will get converted to an electron neutrino is given by
\begin{equation}
P(\nu_\mu \to \nu_e) = \sin^2 2 \theta_G \sin^2 {\pi L \over
\lambda_G},
\label{4}
\end{equation}
where $\lambda_G = {\pi \over E |\phi (r)| \Delta f}$ is the
oscillation length. Although this oscillation mechanism has a
number of similarities with that for neutrinos of non-degenerate
mass, the oscillation length has a markedly different energy
dependence, varying inversely with energy in (\ref{4}), whereas
it is proportional to energy in the massive case \cite{anjan}.
As we shall see, although there exists a fairly wide parameter
space for neutrinos of non-degenerate mass which can explain all
the laboratory experiments, the VEP mechanism for degenerate
mass neutrinos cannot explain all these experiments.

To analyse any neutrino oscillation data, it is useful to divide
the parameter space into three regions: $L<<\lambda$,
$L>>\lambda$ and $L\sim \lambda$.  where $\lambda$ is the
oscillation length. In the large $\lambda$ regime, $\sin^2 2
\theta$ can be as large as unity, and the data constrain
$\lambda$ as a function of the mixing angle $\theta$. As $\theta$
decreases, the maximum excursion for this part of the curve occurs
at approximately $\sin^2 2\theta = \langle P \rangle$, for which
$\lambda = 2L$.  In the
case of the LSND experiment, if one interprets the excess
$\overline{\nu_e}$ events as neutrino oscillations, then the
above oscillation probability would mean both lower and upper
bounds on $\lambda_G$, and hence on $|\phi| \Delta f$.  In the
small $\lambda$ region large numbers of oscillations take place
before the beam reaches the detector, and so the value of
$\sin^2 {\pi L \over \lambda}$ may be assumed to be its average
value $1 \over 2$. In this regime, observation of a small (or
null) neutrino oscillation probability puts both lower and upper
(or just upper) bounds on $\sin^2 2 \theta$ for large $|\phi|
\Delta f$.  In the intermediate region the oscillation length is
comparable to the travel length $L$ and so $\sin^2 {\pi L \over
\lambda}$ varies slowly between $0$ and $1$, putting limits on
$\sin^2 2\theta$ that are quite sensitive to the actual value of
$|\phi|\Delta f$.

For the LSND experiment, the oscillation probability (\ref{4})
in the VEP mechanism can be simplified to
\begin{equation}
P = \sin^2 2 \theta_G \sin^2 (7.62 \times 10^{15} |\phi (r)| \Delta f)
\label{5}
\end{equation}
where we have taken
$\langle L\rangle=30$ meters and the average
neutrino energy is taken to be 50 MeV. The change
on the bounds on $|\phi (r)| \Delta f$ for large $\sin^2 2
\theta_G$ due to the uncertainty in the distance or neutrino
energy are negligible. For $\sin^2 2 \theta_G = 1$, the allowed
region (95\% C.L. from LSND \cite{LSND})
for the violation of the equivalence principle is
\begin{equation}
9.76 \times 10^{-18} > |\phi (r)| \Delta f > 5.03 \times 10^{-18} .
\label{6}
\end{equation}
The small $\lambda$ region occurs when
\begin{equation}
|\phi| \Delta f > 1.35 \times 10^{-16}
\label{7}
\end{equation}
for which the LSND data yields the bound
\begin{equation}
.0029 < \sin^2 2 \theta_G < .011 .
\label{8}
\end{equation}
Thus with the above bounds on the VEP parameter $|\phi|
\Delta f $, one can explain the LSND result.
This would apparently mean that LSND result does not imply a
non-trivial neutrino mass matrix. We shall now demonstrate that
these allowed regions have already been ruled out by other
accelerator experiments.

The E776 experiment at the Brookhaven National Laboratory did
not see any statistically significant excess number of $\nu_e$
($\overline{\nu_e}$) events over the background at a distance of
1 km from the source in a wide-band $\nu_\mu$
($\overline{\nu_\mu}$) beam. Most of the events are above 1 GeV
and peaked around 1.4 GeV. Only 19 events with an expected
background of $25 \pm 5 \pm 3 \pm 3$ were observed and from this
an upper limit on the probability of neutrino oscillations was
determined at 90\% C.L. \cite{E776}.

For large $\sin^2 \theta_G $ this gives a lower limit on the
mass-squared difference consistent with the LSND result for a
neutrino oscillation scenario of neutrinos of non-degenerate
mass \cite{LSND}. However, because of the difference in energy
between the LSND and the E776 experiments, the 1 GeV null result
of the E776 gets translated to
\begin{equation}
|\phi| \Delta f < 3.0 \times 10^{-21}
\label{9}
\end{equation}
in the VEP mechanism. This bound is not consistent with the LSND
result (\ref{6}). In table 1, we present the upper and lower bounds
on $|\phi| \Delta f$ that are permitted within the limits of error
from the E776 and LSND experiments respectively
for several values of $\sin^2 2 \theta_G$.
Over the entire range we find that the upper bound
on the $|\phi| \Delta f$ from the E776 experiment is much less
than the lower bound of $|\phi| \Delta f$ as allowed by the LSND
result. As a result there does not exist any region of the
parameter space of the VEP mechanism which can explain both the
LSND result as well as the E776 experiment for $\sin^2 2
\theta_G > .003$.

\begin{table}[htbp]
\begin{center}
\caption{Bounds on $|\phi(r)| \Delta f$ from E776 (upper) and
LSND (lower)}
\vskip .2in
\begin{tabular}{||c|c|c||}
\hline \hline
&\multicolumn{2}{c||}{}\\
&\multicolumn{2}{c||}{$|\phi(r)| \Delta f$ }\\
\cline{2-3}
$\sin^2 2 \theta_G$&\multicolumn{2}{c||}{}\\
&upper bound from&lower bound from\\
&E776 experiment&LSND experiment\\
\hline \hline
0.002&   8.03E-20&      1.35E-16        \\
0.003& 6.06E-20&    1.02E-16        \\
0.004& 5.09E-20&    8.55E-17        \\
0.005& 4.48E-20&    7.52E-17        \\
0.006& 4.05E-20&    6.8E-17 \\
0.007& 3.72E-20    &6.25E-17        \\
0.008& 3.46E-20&    5.81E-17        \\
0.009& 3.25E-20&    5.46E-17        \\
0.01&  3.08E-20&    5.16E-17        \\
0.02&  2.15E-20&    3.6E-17 \\
0.03&  1.75E-20&    2.93E-17        \\
0.04&  1.51E-20&    2.53E-17        \\
0.05&  1.35E-20&        2.26E-17        \\
0.06&  1.23E-20&    2.06E-17\\
0.07&  1.14E-20&    1.91E-17\\
0.08&  1.06E-20&    1.78E-17\\
0.09&  1.00E-20&    1.68E-17\\
0.1&   9.51E-21&    1.6E-17\\
0.2&   6.71E-21&    1.13E-17\\
0.3&   5.48E-21&    9.19E-18\\
0.4&   4.75E-21&    7.96E-18\\
0.5&   4.24E-21&    7.12E-18\\
0.6&   3.87E-21&    6.5E-18\\
0.7&   3.59E-21&    6.02E-18\\
0.8&   3.35E-21&    5.63E-18\\
0.9&   3.16E-21&    5.31E-18\\
1&     3.00E-21&    5.03E-18\\
\hline \hline
\end{tabular}
\end{center}
\end{table}

We now concentrate on the other regions of the parameter space.
The small $\lambda_G$ region (for which $|\phi| \Delta f$  is so
large that many oscillations occur within the experimental beam
length) sets in for the E776 experiment at
\begin{equation}
|\phi| \Delta f > 8.03 \times 10^{-20}
\label{10}
\end{equation}
which is consistent with (\ref{7}), and implies a bound of
\begin{equation}
\sin^2 2 \theta_G < .003
\label{11}
\end{equation}
on the mixing angle. Thus combining the allowed regions of the
E776 experiment with the LSND result implies that both these
experiments can only be explained by degenerate mass neutrinos
through the VEP mechanism for very large $|\phi| \Delta f > 1.35
\times 10^{-16}$ and for $0.0029 < \sin^2 2 \theta_G < 0.003$.

Even this marginally consistent result may be ruled out as
follows.
If we further argue that gravity couples with matter and
antimatter in the same way (so that gravitational interactions
conserve total lepton number), then we can compare the allowed
parameter space of the VEP mechanism with the $\nu_\mu \to
\nu_e$ oscillation limit as obtained by the SKAT experiment
at Serpukhov \cite{skat}, which provides the most stringent upper
limit on the mixing angle in the small $\lambda$ region
at 90\% C.L. \cite{prd50}.
SKAT measures the ratio of $\nu_e$ to $\nu_\mu$ induced
charged current reactions as observed in a bubble chamber exposed to
a wide band neutrino beam with energies between 3 and 30 GeV and a
neutrino beam length of 270 meters. The small
$\lambda_G$ region sets in for the SKAT average beam energy at
$|\phi| \Delta f > 2.3 \times 10^{-19}$
and the bound on the mixing angle is $\sin^2 2 \theta_G <
.0025$. This fully rules out the the VEP mechanism for
degenerate mass neutrinos.

We can thus conclude that the LSND result cannot be explained by
neutrinos of degenerate mass if other laboratory bounds on
neutrino oscillations are taken into consideration. This result
holds regardless of the gravitational potential at the earth's
surface. In particular, for massless neutrinos there is no
allowed region of the parameter space of the VEP mechanism which
can explain all the experiments.

If we admit the possibility of a non-trivial  neutrino mass
matrix, then all of the above experiments only put bounds on the
VEP parameters.  As discussed in ref. \cite{gasp}, in the
expression for the neutrino oscillation probability in a
two-flavour scenario the oscillation length $\lambda$ is now a
function of $|\phi|\Delta f$, $\Delta m^2$ and two mixing angles,
and the LSND experiment will limit a combination of these
parameters in the large oscillation region via an
appropriate generalization
of eqs. (\ref{5}) and (\ref{6}).
The other experiments will provide further constraints on the
parameter space, and there will be some minimal $\Delta m^2$
which is consistent with all empirical constraints.  We intend
to relate details of this analysis in a forthcoming paper.

In summary, we have shown that the LSND result in conjunction
with other laboratory experiments rules out the possibility of
the VEP mechanism for neutrinos of degenerate mass (this
includes massless neutrinos as well). This situation arises because of the
particular energy dependence of neutrino oscillations in the VEP mechanism in
combination with the differing neutrino energies employed in each experiment.
In the absence of other physical mechanisms for introconversion of neutrino
species, these results imply that neutrinos must have
different nonvanishing masses.  A naturalness
argument would then imply that if the gravitational couplings of
$\nu_e$ and $\nu_\mu$ are the same, then the gravitational
coupling of $\nu_\tau$ should be equal to these, making the VEP
mechanism an unlikely candidate for a neutrino oscillation
mechanism.  Once we admit the possibility that neutrinos are
massive, there is little motivation to consider the VEP
mechanism as the mechanism chiefly responsible for neutrino oscillations.
Present experiments can at best put
bounds on the relevant parameter space.
\vspace{1cm}

\noindent
{\bf Acknowledgements}

This work was supported in part by the Natural Sciences and
Engineering Research Council of Canada. R.B.M. would like to thank
the members of D.A.M.T.P. and of P.R.L. for their hospitality.


\begin{thebibliography}{99}
\baselineskip 16pt
\bibitem{anjan}A. Joshipura, {\it preprint} PRL/94-41 (1994).
\bibitem{solar}R. Davis, D.S. Harmer and K.C. Hoffman, Phys. Rev. Lett
{\bf 20} (1986) 1205; K. Hirata {\it et. al.} Phys. Rev. Lett. {\bf
65} (1990) 1297; {\it ibid>} {\bf 65} (1990) 1301; J.N. Abdurashidov
{\it et. al.}, Phys. Lett. {\bf B328} (1994) 234;
P. Anselmann {\it et. al.} Phys. Lett. {\bf B327} (1994) 377.
\bibitem{solmod}S. Turck-Chieze {\it et. al.}
Phys. Rep. {\bf 230} (1993) 57; J. Bahcall and R.K. Ulrich, Rev. Mod.
Phys. {\bf 60} (1989) 297.
\bibitem{atm}M.C. Goodman, Nucl. Phys. {\bf B38} (Proc. Supp.) (1995)
337; K.S. Hirata {\it et. al.}, Phys. Lett. {\bf B280} (1992)
146; D. Casper {\it et.al.} Phys. Rev. Lett. {\bf 66} (1991) 2561.
\bibitem{barish}B.C. Barish,
Nucl. Phys. {\bf B38} (Proc. Supp.) (1995) 343.
\bibitem{LSND}C. Athanassopoulos {\it et. al.}, LA-UR-95-1238,
nucl-ex-9504002 (1995); B.T. Cleveland {\it et. al.}
Nucl. Phys. {\bf B38} (Proc. Supp.) (1995) 47.
\bibitem{pont}B.M. Ponetcorvo, Sov. Phys. JETP {\bf 34} (1958) 247.
\bibitem{gasp}M. Gasperini, Phys. Rev. {\bf D38} (1988) 2635; {\it
ibid.} {\bf D39} (1989) 3606.
\bibitem{halp}A. Halprin and C.N. Leung, Phys. Rev. Lett. {\bf 67} (1991) 1833;
Nucl. Phys. {\bf B28A} (Proc. Supp.) (1992) 139.
\bibitem{Bah}J.\ N.\ Bahcall, P.\ I.\ Krastev, and C.\ N.\ Leung,
IASSNS-AST 94/54, UDHEP-10-94 (October 1994).
\bibitem{Veps}M.\ N.\ Butler {\it et al.}, Phys.\ Rev.\ {\bf D47},
2615 (1993); A.\ Halprin and C.\ N.\ Leung, Phys.\ Rev.\ Lett. {\bf 67},
1833 (1991); J.\ Pantaleone, A.\ Halprin, and C.\ N.\ Leung, Phys.\ Rev.\
{\bf D47}, R4199 (1993); K. Iida, H. Minakata and O. Yasuda,
Mod. Phys. Lett. {\bf A8} (1993) 1037.
\bibitem{jonas}J. Mureika and R.B. Mann, hep-ph-9501340 (1995).
\bibitem{E776}L. Borodovsky {\it et. al.}, Phys. Rev. Lett. {\bf 68}
(1992) 274.
\bibitem{skat}V.V. Ammosov {\it et. al.}, Z. Phys. {\bf C40} (1988) 487.
\bibitem{prd50}Review of Particle Properties, Phys Rev {\bf D50}, (1994) 1428.
\bibitem{other}
L.A. Ahrens {\it et. al.}, Phys. Rev. {\bf D31} (1985) 2732;
L.S. Durkin {\it et. al.}, Phys. Rev. Lett. {\bf 61} (1988) 1811;
C. Angelini {\it et. al.}, Phys. Lett. {\bf B179} (1986) 307;
G. Zacek {\it et. al.}, Phys. Rev. {\bf D34} (1986) 2621.
\bibitem{Kenyon}
R.J. Hughes, Phys. Rev. {\bf D46} (1992) R2283;
I.R. Kenyon, Phys. Lett. {\bf B237} (1990) 274.

\end{thebibliography}
\end{document}